# Predictive Crystal Plasticity Modeling of Single Crystal Nickel Based on First-Principles Calculations

John D. Shimanek[1], Shipin Qin[1], Shun-Li Shang[1], Zi-Kui Liu[1], and Allison M. Beese[1,2,*]

[1] *Department of Materials Science and Engineering, The Pennsylvania State University, University Park, PA 16802, USA*

[2] *Department of Mechanical Engineering, The Pennsylvania State University, University Park, PA 16802, USA*

*\* Corresponding author: amb961@psu.edu*

**Abstract:**

To reduce reliance on experimental fitting data within the crystal plasticity finite element method (CPFEM), an approached is proposed that integrates first-principles calculations based on density functional theory (DFT) to predict the strain hardening behavior of pure Ni single crystals. Flow resistance was evaluated through the Peierls-Nabarro equation using the ideal shear strength and elastic properties calculated by DFT-based methods, with hardening behavior modeled by imposing strains on supercells in first-principles calculations.  Considered alone, elastic interactions of pure edge dislocations capture hardening behavior for small strains on single slip systems.  For larger strains, hardening is captured through a strain-weighted linear combination of edge and screw flow resistance components.  The rate of combination is not predicted in the present framework, but agreement with experiments through large strains (~0.4) for multiple loading orientations demonstrates a possible route for more predictive crystal plasticity modeling through incorporation of analytical models of mesoscale physics.

## 1. Introduction

Due to their high ductility, plastic deformation in face-centered cubic (fcc) materials has been widely investigated, typically through descriptions of shear stress-strain behavior of the individual slip systems [1]. As both the host element for many superalloys and a prototypical fcc material, pure single crystal Ni is of particular interest, and its mechanical properties have been investigated through both experiments [2–4] and simulations [5,6]. Experimental work on Ni single crystals in the literature has focused on determining the resolved shear stress-strain behavior of their slip systems, including, for example, the initial critical resolved shear stress (CRSS, represented by $\tau_0$ in the present work) and its relationship to orientation and temperature [2].

A fundamental understanding of plasticity based on the evolution of slip system strength has been incorporated into the crystal plasticity finite element method (CPFEM), which has become one of the main computational techniques to relate macroscopic deformation behavior to its slip-based origins [7]. These methods have been used to capture experimentally observed mechanical behavior of both single crystals and polycrystals [8–12]. However, the parameterization of CPFEM models is predominantly accomplished by fitting simulated stress-strain curves to experimental data, underutilizing any physical meaning contained within their parameters and severely limiting their predictive power. Even recent bottom-up approaches to predicting material deformation often start with calibration of slip system behavior to macroscale experimental data [13]. To counter this, some descriptions of slip system strength that depend on physical mechanisms have been developed with the goal of using lower length scale computations to predict these terms [7]. For example, first-principles calculations based on density functional theory (DFT) have been used to predict bcc hardening parameters by considering an isolated mechanism known to be dominant in body-centered cubic (bcc) materials [14]. However, due to the array of complex dislocation interactions responsible for fcc plasticity, first-principles techniques have faced challenges in their extension to predicting hardening parameters in fcc materials [15].

DFT-based first-principles calculations provide a description of atomic processes based on their electronic structures. Advanced techniques have been developed to account more efficiently for far-field strain fields emanating from dislocations while maintaining atomistic accuracy near the core [14,16,17], but complex interactions of large numbers of dislocations remain out of reach of first-principles methods. Consequently, the present work makes no attempt to explicitly consider dislocations and instead focuses on the properties of ideal shear strength and elasticity, which can be related to physical parameters in CPFEM models and can be reliably obtained by DFT-based first principles calculations through the imposition of strains [18,19].

The present work introduces a method of linking a computationally tractable problem, the ideal shearing process, to a realistic description of macroscopic deformation, combining the utility of CPFEM modeling with the predictiveness of DFT-based calculations. In this proposed approach, the effects of the elastic field on ideal shear strength due to long-range interactions between dislocations are mimicked by applying pre-strains in the first-principles calculations. The flow resistance for pure edge and pure screw dislocations was then predicted using the Peierls-Nabarro model [20,21]. In contrast to CPFEM frameworks with numerous parameters whose physical interpretations are unused during fitting to experimental data, two of only three hardening model parameters were predicted based on their physical analogue in the context of first-principles methods. The parameterized CPFEM model was used to predict the macroscopic stress-strain curves of various single crystal tensile tests to small tensile strains. Due to the importance of screw characteristics of dislocations at high dislocation densities [22], the present work proposes to model the flow stress, which is used to parameterize an established CPFEM hardening model, as a linear combination of flow stresses of edge and screw dislocations weighted by plastic strain. The linear coefficient presently relies on a fitting procedure, but the framework allows for the use of first-principles results while also considering mesoscale physics, the incorporation of which will be the subject of future work. The results of CPFEM simulations were compared with available literature data on the strain hardening behavior of single crystal Ni of multiple orientations at large strains.

## 2. Methodology

### 2.1 Approach

In a single crystal fcc material, dislocations themselves are the main obstacles that inhibit dislocation movement and thus the major strain hardening mechanism [22,23]. In the small strain range, the dislocation density is relatively low, and the dislocation interaction primarily occurs through long-range elastic fields [1]. In the large strain range, short-range dislocation interactions become the major strain hardening source as the dislocation density is high and the dislocation mean free path is low [22,24]. In short-range dislocation interactions, dislocation cores make contact with each other to form jogs or junctions. Because junctions formed by dislocation core reactions can be sessile, junctions are considered to be the major source of strain hardening in stage II deformation of fcc crystals [22]. The model predictions considering only junctions have shown satisfactory agreement with experiments in the literature [22,25,26].

In DFT-based calculations, direct consideration of dislocations is challenging due to the high computational cost of the calculations, which limits their size, and the inherently extended nature of dislocations. Therefore, explicit first-principles calculations even of single dislocations have been made only with the help of elastic Green function solutions to account for the far-field elastic distortions, attenuating image forces due to periodic boundary conditions and allowing the accurate yet expensive DFT-based calculations to relax only those atoms deemed to be part of the dislocation core [14,17,27,28].

In the present work, a different approach is proposed to consider dislocations in an indirect manner. Specifically, the method adopted here relies on the improved analytic form to estimate Peierls stress proposed by Joós et al. [29], which is based on elastic properties and the ideal shear strength along the partial slip system and is widely employed [30–33]. Both the elastic constants and the ideal shear strength can be calculated through DFT-based methods, and, critically, these values can be calculated even when the crystal structure is already under the influence of an orthogonal shear strain. In the present work, the resulting increase in the ideal shear strength in these "pre-strained" structures is interpreted as the general influence of elastic fields on flow resistance and is used as a proxy for hardening due to long-range interactions of dislocations. The approach is described in general by a schematic in Figure 1 then through a discussion in the following two sections, with further calculational details in the Supplemental Materials.

The central postulation of the present work is that the response of the ideal crystal to elastic strains contains information relevant to a description of macroscopic deformation. Strain applied to the ideal crystal in one direction is used to determine, through the Peierls-Nabarro equation, the stress required to move a single dislocation in an otherwise perfect and strain-free lattice. Adding strain in an orthogonal in-plane direction increases the difficulty of the ideal shear process, analogously to the way that strain field interactions of multiple dislocations increase the difficulty of their motion through a realistic crystal. As will be discussed in Section 3.5, the above procedure predicts insufficient hardening at large strains, where the changing nature of the dislocation network must be taken into account. To this end, a model is proposed combining the effects of edge and screw dislocations as a function of strain to obtain new hardening parameters suitable for large strain predictions. While the model for large strains does not explicitly describe pinning effects or hardening due to forest dislocations, it generally considers the evolution of the most mobile dislocation segments as these segments increasingly become exhausted through formation of more sessile junction segments. The overall strength of the slip system is then tied to the mobility of its most mobile segments, as it must be for plastic strain to be accomplished through slip.

**2.2 Crystal plasticity model**

The crystal plasticity framework presented by Huang [34] is adopted in the current work. In this framework, strain hardening is described as the evolution of the CRSS on one slip system due to the shear strain on any slip system:

$$\dot{\tau}_c^\alpha = \sum_{\beta=1}^{l} h_{\alpha\beta} |\dot{\gamma}^\beta| \qquad \text{Eq. 1}$$

where $\tau_c^\alpha$ is the CRSS on slip system $\alpha$, $\gamma^\beta$ is the shear strain on slip system $\beta$, and $h_{\alpha\beta}$ is the hardening matrix. A form of $h_{\alpha\beta}$ presented by Peirce et al. [35] is adopted in the present work for the simplicity of its form and the interpretability of the individual parameters between length scales. Peirce et al. [35] proposed that:

$$h_{\alpha\beta} = q_{\alpha\beta} \left[ h_0 \operatorname{sech}^2 \left| \frac{h_0 \gamma}{\tau_s - \tau_0} \right| \right] \qquad \text{Eq. 2}$$

where

$$q_{\alpha\beta} = \begin{cases} 1, & \alpha = \beta \\ 1.4, & \alpha \neq \beta \end{cases} \qquad \text{Eq. 3}$$

characterizes the difference between self-hardening ($\alpha = \beta$) and latent hardening ($\alpha \neq \beta$), for which a ratio of 1.4 is widely accepted in the literature [7,36]. With this form, the slip system level strain hardening curve increases monotonically with a decreasing slope and approaches a saturation value asymptotically. The initial slope of this curve is controlled by $h_0$, the saturation value is controlled by $\tau_s$, and the initial CRSS value is $\tau_0$.

It should be noted that even though less physically-motivated than the model of Taylor based on dislocation density [37], the hardening model by Peirce et al. [35] does not rely on explicit descriptions of dislocations that would be prohibitive in first-principles methods due to the computational expense. However, the models of Taylor [37] and Peirce et al. [35] describe the same deformation response and therefore must also describe the effects of collective dislocation motion, whether explicitly or implicitly. Similarly, other forms of $h_{\alpha\beta}$, discussed in Ref. [7], describe deformation using terms specific to dislocation motion such as interaction strength, lock formation, dipole formation, and annihilation processes, each contributing at least one additional fitting parameter to the overall hardening law for a total parameter set that can easily number over 15.

The hardening modulus adopted in the present work (Eq. 2) has found application to the deformation behavior of single crystal tungsten [38,39], single crystal copper [40], polycrystal copper [41], friction stir welded aluminum [42], Ti-6Al-4V [43], and dual-phase steels [44]. The deformation modes to which it has been applied range from nanoindentation [38–40,43] to wire tension tests [41] to the deformation of representative volume elements that approximate a microstructure's bulk mechanical response [42,44]. In each case, model parameters were determined by fitting to experimental stress-strain or load-displacement data. The key novelty and strength of the present study is that $\tau_0$ and $h_0$, as well as all elastic constants, were predicted through DFT-based computations and thus were not fit using experimental data. The value for $\tau_s$ was taken from results reported in the literature.

**2.3 First-principles calculations of flow resistance**

The initial CRSS, $\tau_0$, is the minimum stress required to initiate plastic deformation [18], which for perfect crystals corresponds to the ideal shear strength, $\tau_{IS}$, while more generally this corresponds to the initial flow resistance, $\tau_f$. In the present work, $\tau_f$ is estimated using the Peierls-Nabarro model to find the Peierls stress, $\tau_p$, the minimum stress required to move a dislocation [20,21]. For the spatially extended strain fields surrounding dislocations, common to pure metals, the Peierls-Nabarro equation is given in Eq. 4 [29,45].

$$\tau_P = \frac{Kb}{a} \exp(-2\pi\zeta/a) \approx \tau_f \qquad \text{Eq. 4}$$

Here, $b$ is the Burgers vector, $a$ is the row spacing of atoms within the slip plane (for example, $a = a_0\sqrt{6}/4$, where $a_0$ is the lattice parameter, for the case of $\{111\}\langle11\bar{2}\rangle$ shear deformation of an fcc lattice), and $\zeta$ is the half-width of the dislocation, given as:

$$\zeta = \frac{Kb}{4\pi\tau_{IS}} \qquad \text{Eq. 5}$$

The elastic factor, $K$, is direction-dependent for an anisotropic crystal like pure Ni. Analytical forms for anisotropic elastic factors have been derived, e.g., in Ref. [46], and depend on the character of the dislocation, with variants existing for both pure edge and pure screw dislocations. Their full forms are given in the Supplementary Material.

The ideal shear strength in Eq. 5 can be predicted directly by pure alias shear – a deformation mode more representative of the slip process than affine shear [19,47,48]. Alias shear involves only one sliding layer, with the atoms in other layers initially remaining in their original positions [19,47,48]; see Figure 2b. The relaxations of a pure alias shear include all degrees of freedom of a supercell except for the fixed shear angle as well as any other imposed constraints, such as the pre-strain deformation discussed below.

A 6-atom orthorhombic supercell, shown in Figure 2a, was adopted for fcc Ni with its lattice vectors $a_{orth}$, $b_{orth}$, and $c_{orth}$ of lengths $0.5\sqrt{6}a_0$, $0.5\sqrt{2}a_0$, and $\sqrt{3}a_0$ (where $a_0$ is the conventional lattice parameter of fcc Ni) and parallel to the $[11\bar{2}]$, $[\bar{1}10]$, and $[111]$ directions, respectively. This reorientation of the standard unit cell allows for shear in the $[11\bar{2}]$ direction within the $(111)$ plane, the softest direction of the perfect crystal and the slip system of partial dislocations [19,46,48], to be accomplished by controlling the shear angle between the 1$^{st}$ and 3$^{rd}$ lattice vectors of the supercell. Pure alias shear of increasing magnitude along this direction of the crystal allows for calculations of the ideal shear strength, i.e., the maximum shear stress that the crystal can withstand before elastic destabilization. Similarly, pre-strains along $[\bar{1}10]$ within the same plane are accomplished by constraining the angle between the 2$^{nd}$ and 3$^{rd}$ lattice vectors. Strains in the $(111)[\bar{1}10]$ system are held constant throughout the calculations of the ideal shear strength along $[11\bar{2}]$ and serve as a method of probing the dependence of ideal shear strength on an imposed elastic field. As discussed in Section 2.1, pre-strains on orthogonal slip systems represent the influence of one slip system on another through their elastic fields and will become important in predicting a strain hardening response.

Elastic properties can be predicted by computing stresses under given strains by means of first-principles calculations and Hooke's law, with imposed non-zero strains being ±0.007 and ±0.013, as previously described [49,50]. The elastic factor for both edge and screw dislocations can be calculated and applied to the Peierls-Nabarro equation (Eq. 4) for evaluation of the flow resistance as a function of strain and dislocation character. Further details of the first-principles calculations, for both elastic properties and flow resistance, are given in the Supplementary Material.

## 3. Results and discussion

### 3.1 Results from first-principles calculations

Investigation into the effect of the number of {111} layers contained in the supercell showed that 3 atomic layers, which is 12 atoms and the minimum needed for this deformation mode, produced the highest ideal shear strength values. The ideal shear strength of pure Ni found in the present study, 5.15 GPa, matches well with the estimated values in the literature, which range from 5 to 8 GPa [19,51,52]. The nature of the layer dependence and literature comparisons are discussed in more detail in the Supplementary Material.

The ideal shear strengths under various pre-strains were calculated and shown in Table 1. It can be seen that the ideal shear strength increases with the magnitude of the orthogonal pre-strain while the shear strain at which the ideal strength is reached decreases. As mentioned in Section 2.1, the orthogonal pre-strain can be interpreted as the effect of the elastic field of one slip system on the deformation behavior of another. Therefore, the increase in ideal shear strength of one slip system as a function of the shear strain on another is indicative of strain hardening behavior, the quantification of which is discussed in the upcoming Section 3.2.

The elastic constants of fcc Ni in terms of the 6-atom orthorhombic cell ($c'_{ij,\text{orth}}$) are summarized in Table 2. Note that by adopting the relationship given by Hirth and Lothe [46], $c'_{ij,\text{orth}}$ can be transformed to $c_{ij,\text{cub}}$, which are the elastic constants in terms of the 4-atom conventional cubic cell for comparison with experimental data. The predictions without pre-strain agree with the experimental elastic constants extrapolated to 0 K [53]. With the elastic constants and the ideal shear strengths established as functions of pre-strain, flow resistance can be calculated through the Peierls-Nabarro framework (Eq. 4) at each pre-strain. The elastic factor, in turn, depends upon the character of the dislocation, with the limiting cases of pure edge and pure screw given in the Supplementary Material. Note that the elastic factor at each pre-strain was calculated based on the elastic constants of the pre-strained structure to better capture the local elastic environment so that each input value to Eq. 4 comes from a calculation using the same initial (pre-strained) structure.

The predicted flow resistance ($\tau_f$ or $\tau_P$) at 0 K are compared with experimental $\tau_0$ values at room temperature in Table 1 [2,3,54,55]. See the Supplementary Material for a discussion of the appropriateness of comparing 0 K predictions with room temperature measurements. The predicted $\tau_f$ values of edge dislocations (9.4 MPa without pre-strain) agree well with experimental $\tau_0$ values (5.5 to 19.6 MPa [2,3,54–60]), but those of screw dislocations (117.7 to 308.7 MPa without pre-strain) are significantly higher. It can be seen that the predicted $\tau_f$ value increases with the magnitude of the pre-strain. For example, $\tau_f^{edge}$ increased from 9.4 to 11.1 MPa (an increase of about 18 %) as the magnitude of the pre-strain increased from 0 to 0.049. The increase of $\tau_f$ stems mainly from the increase of $\tau_{IS}$ rather than the elastic properties (see Eq. 4 and values in Table 1).

**3.2 CPFEM model parameters from first-principles calculations**

For small strains, when an ample portion of highly-mobile edge-type segments exist in the dislocation network, the most meaningful flow resistances are those calculated through the Peierls-Nabarro equation using elastic factors for pure edge dislocations. The high mobility of edge-type dislocations is seen not only in the initial flow resistance values of Table 1 but also in the relative ease with which edge dislocations break free from the junctions formed during strain hardening [61]. Therefore, the predictions based on pure edge and pure screw dislocations must be combined at large deformations as the highly-mobile segments are exhausted, leaving behind junctions and segments of an increasingly screw character. The procedure used to combine the flow resistances for both dislocation types as a function of strain will be discussed in Section 3.5. In the present section, the procedure for quantifying hardening behavior based on first-principles calculations, the critical length-spanning translation step of the present work, will be given in the context of small strains.

As discussed in Section 2, DFT-based calculations predicted the flow resistance of a dislocation gliding along slip system $\alpha$ under the influence of an elastic field from other dislocations. The intensity of the elastic field can be mimicked by the pre-strain imposed in the DFT-based calculations. This pre-strain corresponds to the local effect of the shear strain on a latent slip system caused by the long-range elastic field of dislocations and is the $\gamma^\beta$ ($\alpha \neq \beta$) in Section 2.2. Note that while $\gamma^\beta$ represents a plastic strain, it is also indicative of the slip system activity, which results in the generation and interaction of dislocations. In a Taylor-like model, this slip system activity information might be encoded into a dislocation density parameter for each slip system. Since long-range dislocation interactions are conveyed through elastic strain fields, the results from first-principles calculations in the present work, describing the effect of elastic strain on flow resistance, is applicable to the mesoscale description of slip defined in the CPFEM hardening equations. By imposing different levels of pre-strain, the relationship between $\tau_c^\alpha$ and $\gamma^\beta$ was predicted (i.e., $\tau_f$ versus the pre-strain $\gamma_{110}$ in Table 1). In this case, Eq. 1 through Eq. 3 can be simplified as:

$$\dot{\tau}_c^\alpha = 1.4 \left[ h_0 \operatorname{sech}^2 \left| \frac{h_0 \gamma^\beta}{\tau_s - \tau_0} \right| \right] |\dot{\gamma}^\beta|, (\alpha \neq \beta) \qquad \text{Eq. 6}$$

where $h_0$, $\tau_0$, and $\tau_s$ are model parameters. By matching the relationship between $\tau_c^\alpha$ and $\gamma^\beta$ determined from Eq. 6 (note that $\tau_c^\alpha = \tau_0$ when $\gamma^\beta = 0$) with that predicted in DFT-based calculations, the values of $\tau_0$ and $h_0$ were determined. As shown in Figure 3, for small strains, the hardening relation is approximately linear, and $\tau_0$ and $h_0$ correspond to the intercept and slope, respectively. Note that because DFT-based predictions are limited to small strains, where $h_0$ and $\tau_0$ play a dominant role, a value reported in the literature was adopted for $\tau_s$ (40 MPa [5]). The determined parameter values are summarized in Table 3.

### 3.3 Experimental results in the literature

To show the predictive accuracy of the present approach, geometrical models of Ni single crystal tensile tests from the literature were constructed and combined with the predicted hardening parameters summarized in Table 3 to evaluate macroscopic stress-strain responses. The experiments considered in the present work include two uniaxial tension tests reported by Haasen [2] on 99.999% purity Ni wire specimens with a diameter of 2.24 mm and a length of 71.12 mm, and a uniaxial tension test reported by Yao et al. [60] on 99.999% purity Ni specimens with a gauge section size of $2.5 \times 5.5 \times 0.25\ mm^3$, both for single crystals. Figure 4a provides the resolved shear stress-strain curves reported in these publications [2,60]. The process of calculating engineering values from the resolved shear stress-strain curves is detailed in the Supplementary Material, while the final engineering stress-strain curves are shown in Figure 4b. Note that the loading directions with respect to the crystal orientation are different for each test, i.e., $\langle\bar{1}\,5\,10\rangle$ and $\langle\bar{1}28\rangle$ by Haasen [2], and $\langle 011\rangle$ by Yao et al. [60].

Discrepancies in the reported literature on pure Ni single crystal CRSS and flow behavior stem from differences in material purity, initial dislocation density, and potential experimental uncertainties. A method must therefore be adopted to evaluate these differences so that they may be considered when comparing computational results to experimental data. Here, differences in experimental results were evaluated by comparing their initial CRSS values, which are independent of the assumptions adopted for converting force-displacement data to resolved shear-stress strain data. Supplementary Figure S2 shows the initial CRSS value of pure Ni reported by ten different groups [2,3,54–60,62]. Since the value reported by Latanision et al. [62] is significantly higher than the other reported values, it was excluded from evaluation in the present study. The rest of the experimental data all lie between 5 MPa and 20 MPa, and the statistics of these data are shown in Supplementary Table S2. According to the statistical analysis of the initial CRSS reported by nine different groups over more than 80 years, the experimental data in the literature exhibited a relative error of 43%.

**3.4 DFT-based CPFEM predictions at small strains**

To simulate the tests reported in the literature, the full geometry of the specimens in each test was modeled. All of the specimens were discretized with 0.2 mm hexahedral full integration elements (element type C3D8 [63]) in the gauge region, and the models contain 20,590 elements for the wire specimen by Haasen [2] and 2,176 elements for the dogbone specimen by Yao et al. [60]. In both models, the vertical movement of the bottom nodes was constrained while a uniform vertical displacement was applied to the top nodes. The horizontal movements of all top and bottom nodes of the flat dogbone specimen in Yao et al.'s study were also constrained to avoid potential out-of-plane distortion [64]. The crystal plasticity model was implemented in the commercial finite element software ABAQUS through a user subroutine UMAT [63] originally developed by Huang [34,65].

The simulated engineering stress-strain curves compared to the respective experimental results are shown in Figure 5, where it can be seen that the initial yield stresses in all of the tests were reasonably predicted. Table 4 provides a detailed comparison between experimental and predicted initial yield stresses for all tests. The model to obtain parameters for the CPFEM calculations, whose stress-strain results are shown, only considered elastic interactions of pure edge dislocations, and the agreement on yield stress between experiment and model supports the premise that edge dislocations are initially dominant at low strains due to their higher mobility.

While both $[\bar{1}\ 2\ 8]$ and $[\bar{1}\ 5\ 10]$ loading orientations result in single slip, the latter is slightly further from the angle at which a second, conjugate slip system becomes active during uniaxial tension according to the geometrical effects considered in Schmid's law. Among the experimental stress-strain data shown in Figure 5, the low hardening regime characteristic of single system slip is most clearly seen for the case of the $[\bar{1}\ 5\ 10]$ orientation results, with which the hardening prediction shows good agreement up to strains of 0.06. Model agreement in this low-hardening region suggests that the first-principles methods effectively emulate long-range elastic interactions between edge dislocations within the same slip system. The underprediction of hardening for larger strains, and for loading orientations falling nearer to multi-slip conditions ($[011]$ is expected to begin immediately with four active slip systems) indicates that capturing large strain hardening behavior requires the consideration of different mechanisms, including junction formation and the overall mobility evolution of the dislocation network. Note that using elastic pre-strains to mimic the long-range elastic interactions between dislocations generated by plastic strain represents a limiting case, overestimating the elastic strain on the mobile dislocation due to the overall macroscopic plastic strain. As a result, the hardening rates in Figure 5 represent upper bounds to their small-strain estimation within this framework, further supporting the need to consider additional mechanisms.

**3.5 Modeling and predictions at large strains**

At large strains, as the highest mobility dislocation segments become exhausted, the effect of dislocation segments of both edge and screw character must be considered. Dislocations come into contact and form junctions that often exhibit screw character [66–69], which are a major contributor to the strain hardening of fcc crystals in the large strain range [22]. This indicates that the relative contribution to strain hardening from screw dislocations, and other segments that are difficult to move by an applied stress, increases with plastic strain. Therefore, in the present study, the following model is proposed to account for the increasing influence of screw components to strain hardening with plastic strain:

$$\tau_c^{\alpha,es} = (1 - w\gamma^\beta)\tau_f^{edge} + w\gamma^\beta \tau_f^{screw} \qquad \text{Eq. 7}$$

where $w$ is a weighting factor that controls the contribution from each type of dislocation, $\gamma^\beta$ is the shear strain on slip system $\beta$, $\tau_f^{edge}$ and $\tau_f^{screw}$ are the predicted CRSS in the DFT-based calculations (see Eq. 4) for pure edge and pure screw dislocations, respectively, and $\tau_c^{\alpha,es}$ is the CRSS on slip system $\alpha$ (see Eq. 1) considering contributions to the strain hardening from both edge and screw dislocations. Both $\tau_f^{edge}$ and $\tau_f^{screw}$ increase with $\gamma^\beta$, as shown in Figure 3, and their combination through Eq. 7 is shown schematically in Figure 6. By including $\gamma^\beta$ in the model, the influence of both types of dislocations are included naturally: (1) edge dislocations are dominant at small strains ($\tau_c^{\alpha,es} = \tau_f^{edge}$ when $\gamma^\beta = 0$), and (2) the influence of screw dislocations increases with increasing strain, in accordance with the studies of Kubin et al. [70], Guruprasad and Benzerga [71] and Huang et al. [72].

The weighting factor $w$ in Eq. 7 was adjusted to reproduce the stress-strain curve by Yao et al. [60]. Figure 7 shows that the CPFEM simulations agreed well with experiments over the full experimental strain range with $w = 0.33$. Note that Eqn. 7 produced a new resolved shear stress-strain curve, based on which a new set of $\tau_0$ and $h_0$ values were determined. The newly-found parameters belong to the same established hardening framework outlined in Eqns. 1-3. Specifically, $h_0$ is a function of the weighting factor, $w$, and its value reflects the contributions from both edge dislocations and screw dislocations based on the slope of the relationship given in Eq. 7. The new parameter values are summarized in Table 3. As discussed in Section 3.2, the saturation stress ($\tau_s$ in Eq. 2) cannot be determined from DFT-based calculations; therefore, $\tau_s$ was calibrated to be 300 MPa based on the experimental data in Figure 7b. Note that the value of $\tau_s$ only affects the stress-strain curve in the large strain range. In the present study, one simulation with $\tau_s$ being an order of magnitude higher than 300 MPa resulted in a stress-strain curve that was only slightly different for engineering strains greater than 0.6. Therefore, the excellent agreement in Figure 7 is primarily attributed to the value of $h_0$, which is derived from the DFT-based calculations and the weighting factor $w$. While the weighting factor for the present calculations was fit to one of the single crystal Ni curves, its physical meaning as the rate of average character evolution of a dislocation network allows for its prediction based on other types of simulations that explicitly consider representative numbers of dislocations. Such mesoscale investigations are beyond the scope of the present work; here, results of the large strain parameterization are used to demonstrate how first-principles results may be incorporated into CPFEM calculations.

The wire tension tests performed by Haasen [2] were simulated again using the newly determined parameters that consider the influence of both edge and screw dislocations. Figure 8 shows that in the new CPFEM simulations of Haasen's tests, the flow stress agrees reasonably well with the experimental results up to large strains, with the largest discrepancies at the beginning of the $[\bar{1}\ 5\ 10]$ deformation (where the overprediction is 17 MPa, or 68%) and in the middle of the $[\bar{1}\ 2\ 8]$ deformation simulation (which underpredicts by 26 MPa, or 20%, at worst). Overall, the mean absolute percentage error, calculated as the average of the relative differences between predicted and experimental stress values for each experimental strain value, is 11% for the $[\bar{1}\ 2\ 8]$ loading and 13% for the $[\bar{1}\ 5\ 10]$ loading. The mean absolute error, where the absolute stress differences were again calculated at each experimental strain value and averaged, is 13 MPa for $[\bar{1}\ 2\ 8]$ and 7 MPa for $[\bar{1}\ 5\ 10]$.

The present model relies on the dominance of partial slip systems and therefore has natural limitations, prohibiting its direct use in bcc materials or those where dislocations are unlikely to dissociate into partials. Even though forms of the Peierls-Nabarro equation exist for more compact dislocation cores [29], their use within the present framework for more complex crystal structures could be complicated by the existence of specific dislocation mechanisms that dominate plastic flow, as in the case of kink nucleation in bcc materials [14]. The focus of the present work is the fcc case of pure Ni, and it should be emphasized that in the above calculations, only the weighting factor, $w$, and the saturation stress, $\tau_s$ (whose contribution to the accuracy of the predictions was negligible), were fitted from a macroscopic stress-strain curve, while all other parameters were predicted from DFT-based calculations. In contrast, existing physics-based crystal plasticity models in the literature generally feature large numbers of fitting parameters, with the fitting process in practice diminishing the physical significance of each parameter.

## 4. Conclusions

In the present work, an approach has been developed to predict the macroscopic stress-strain behavior of pure Ni single crystal. Instead of calibrating CPFEM model parameters solely using macroscopic experimental results, a standard practice in the literature, the present CPFEM simulations employed DFT-based first-principles calculations of flow resistance at 0 K in terms of the predicted ideal shear strength and elastic properties. The key findings of the present work are:

- Initial values of the flow resistance, calculated based on the Peierls-Nabarro equation with pure edge dislocation elastic factors at 0 K, matched well with experimental critical resolved shear stress values at room temperature. This agrees with established theory on the high mobility of edge dislocations in fcc materials.

- Through the application of increasing orthogonal in-plane shear strains, the increasing ideal shear stress along the main strain direction was predicted. This elastic pre-strain in a perfect crystal is analogous to the strengthening effects of long-range strain fields between dislocations and serves as a linkage between atomistic and continuum descriptions of hardening.

- The increase in flow resistance with pre-strain based on ideal shear strengths and pure edge elastic factors gave a description of small-strain hardening behavior. When these first-principles results were incorporated into an established hardening model, the CPFEM single crystal stress-strain predictions agreed with experimental results for yield and, if present, for the initial, low-hardening regime. At larger strains, the underprediction of the hardening rate evidenced the need to incorporate strengthening effects from screw dislocation segments into the hardening model.

- A simple model for CPFEM parameters is proposed that combines the contributions to flow resistance from edge and screw dislocations as a function of strain, in agreement with the concept that junctions and less-mobile dislocation segments become important after plastic deformation is initiated by the motion of edge dislocation segments. The strength evolution behavior resulting from the combination of the edge and screw contributions is easily incorporated into CPFEM, as it can be described by the same hardening model with modified parameters. The physical interpretation of the rate of combination allows for future cooperation with mesoscale modeling to make the large strain modeling fully predictive.

- With the combination of edge and screw contributions to strength, the present work accurately captures the strain hardening of Ni single crystals of various orientations through large deformations.

**Acknowledgements**

This work was financially supported by the U. S. Department of Energy (DOE) via award no. DE-FE0031553 and the Office of Naval Research (ONR) via contract no. N00014-17-1-2567. First-principles calculations were carried out partially on The Pennsylvania State University's Institute for Computational and Data Sciences' Roar supercomputer, partially on the resources of NERSC supported by the DOE Office of Science under contract no. DE-AC02-05CH11231, and partially on the resources of XSEDE supported by NSF via grant no. ACI-1548562. JDS was supported by the Department of Energy National Nuclear Security Administration Stewardship Science Graduate Fellowship, provided under cooperative agreement number DE-NA0003960.

**Data Availability**

All relevant data are available from the authors.

**Conflict of Interest**

On behalf of all authors, the corresponding author states that there is no conflict of interest.

**Figures**

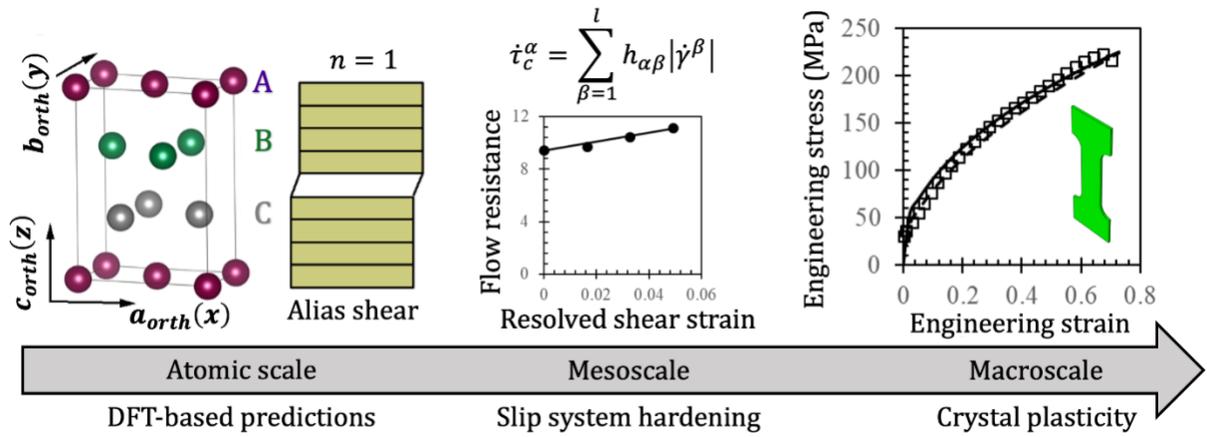

**Figure 1:** A schematic of the overall approach proposed in the current work, showing the transfer of information from the atomic scale ideal shear process to a mesoscale description of hardening on a slip system level to, finally, a description of macroscale deformation of single crystal samples.

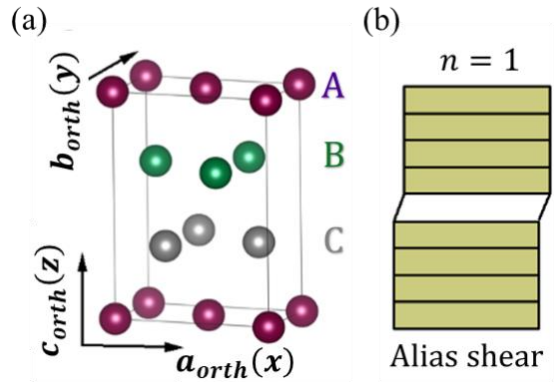

**Figure 2:** (a) Three-layer six-atom orthorhombic cell of fcc lattice with its lattice vectors $a_{orth}$ ($x$), $b_{orth}$ ($y$), and $c_{orth}$ ($z$) parallel to the $[11\bar{2}]$, $[\bar{1}10]$, and $[111]$ directions of the conventional fcc lattice; where the letters A, B, and C indicate three closed packed (111) planes. (b) Schematic diagrams of alias shear demonstrated using two cells in the shaded area, with atoms in only one plane involved in shear (i.e., the number of involved atomic planes, $n$, is one, shown as the unshaded area).

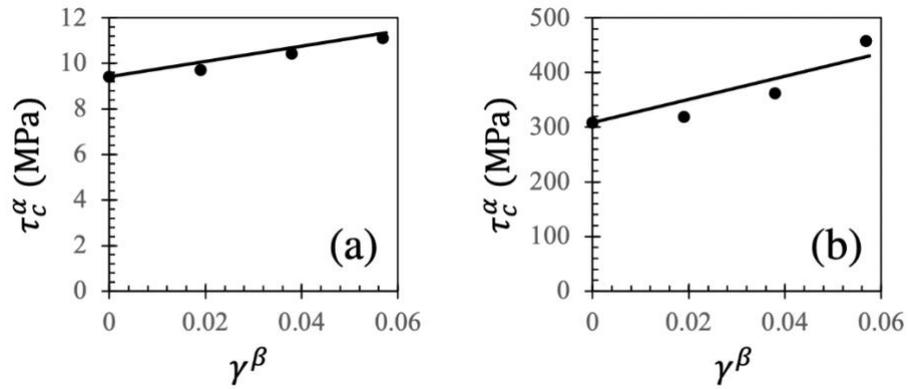

**Figure 3:** Critical resolved shear stress on slip system $\alpha$ as a function of shear strain on slip system $\beta$ where the dislocation character of the $\alpha$ system is taken to be (a) edge or (b) screw. Symbols represent DFT-based predictions, and lines show the corresponding CPFEM model curves.

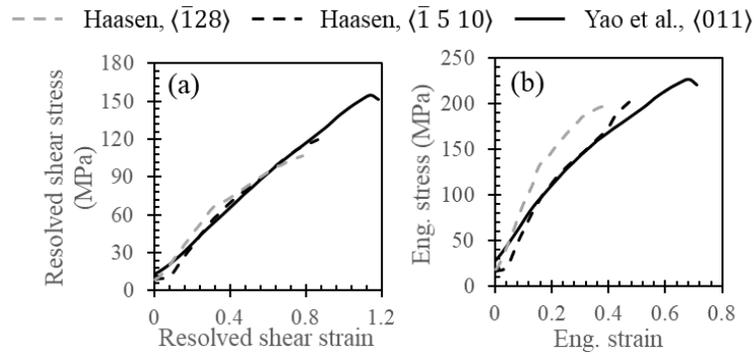

**Figure 4:** (a) Resolved shear stress vs. resolved shear strain and (b) engineering stress vs. engineering strain for pure Ni bulk single crystals in literature [2,60]. The crystallographic directions in the legend indicate the loading direction during the tests.

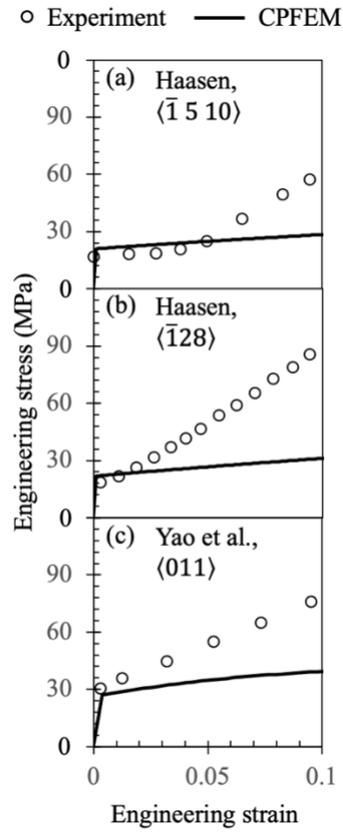

**Figure 5:** CPFEM predictions for (a,b) Haasen's tests [2] and (c) Yao et al.'s test [60] from edge dislocation based flow resistance compared to experimental results (symbols) by Haasen et al. [2] and Yao et al. [60].

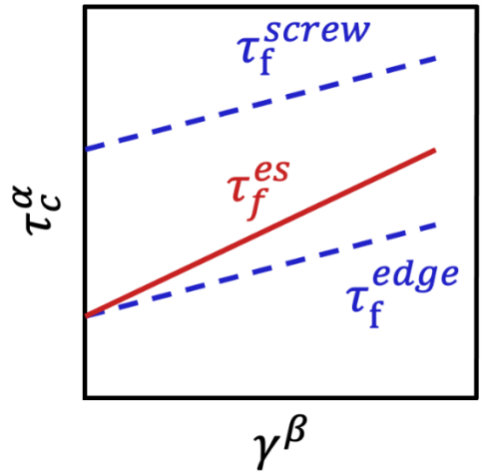

**Figure 6:** Schematic of the combination procedure described by Eq. 7 in which a fraction of the screw component flow stress is added to the edge component flow stress as a function of strain.

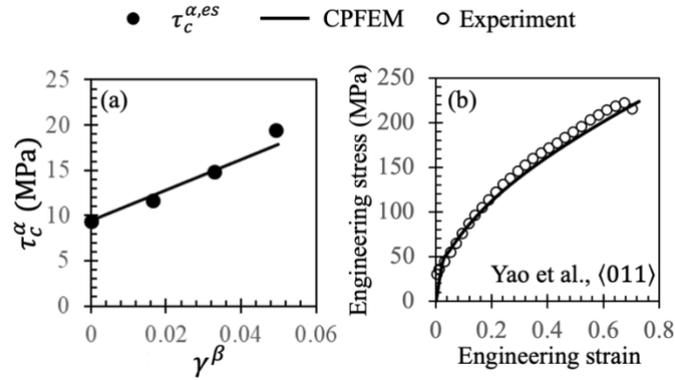

**Figure 7:** (a) Flow resistance, $\tau_c^{\alpha,es}$, on slip system $\alpha$ that combines contributions from both edge and screw dislocations as a function of shear strain on slip system $\beta$. The calculated flow resistances are shown as symbols, and the corresponding CPFEM fits are also shown (lines). (b) CPFEM simulated engineering stress-strain curves (lines) of Yao et al.'s test [60] and the corresponding experimental results (symbols).

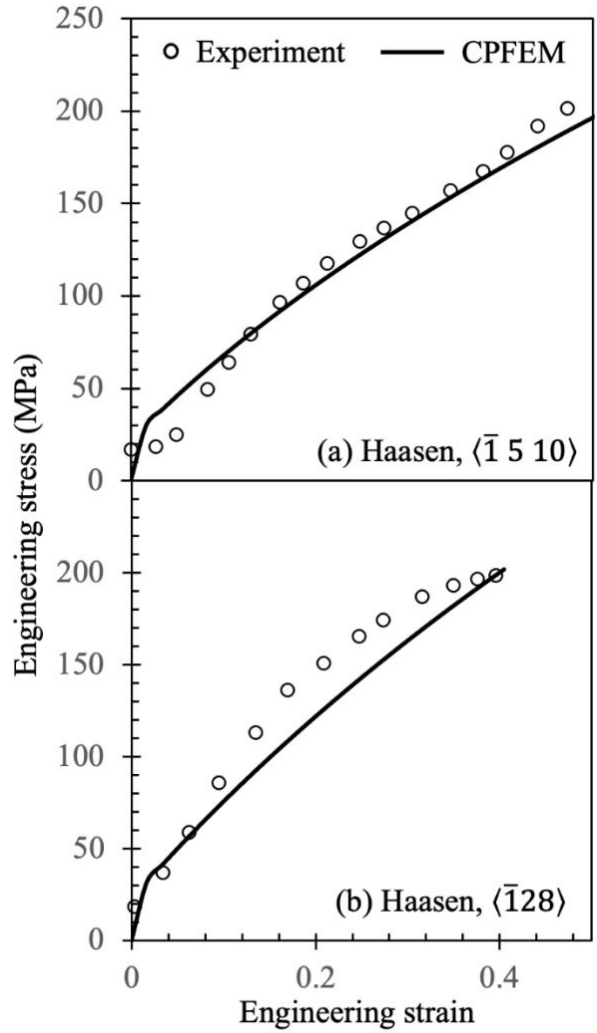

**Figure 8:** Engineering stress-strain curves for experiments (symbols) along (a) $\langle \bar{1}\,5\,10 \rangle$ and (b) $\langle \bar{1}28 \rangle$ in ref. [2] compared to CPFEM predictions from the present study (lines) for the combined edge and screw predictions by DFT-based calculations (see Eq. 7).

## Tables

**Table 1:** Ideal shear strength ($\tau_{IS}$) of fcc Ni due to pure alias shear of $\{111\}\langle11\bar{2}\rangle$ with pre-strain $\gamma_{110}$ along the $[\bar{1}10]$ direction, together with the predicted flow resistance ($\tau_f$, MPa) at 0 K for edge and screw dislocations in comparison with experimental CRSS values ($\tau_{CRSS}$, MPa) at room temperature.

| Properties | $\gamma_{110} = 0.000$ | $\gamma_{110} = 0.016$ | $\gamma_{110} = 0.033$ | $\gamma_{110} = 0.049$ |
|---|---|---|---|---|
| $\gamma_{112}$ [a] | 0.128 (0.780) | 0.126 (0.770) | 0.124 (0.754) | 0.120 (0.732) |
| $\tau_{IS}$ (in GPa) | 5.15 | 5.16 | 5.19 | 5.26 |
| $\tau_f^{edge}$ [b] | 9.4 | 9.7 | 10.4 | 11.1 |
| $\tau_f^{screw}$ [b] | 308.7 | 318.6 | 362.1 | 457.5 |
| $\tau_0$ (Expt.) | 5.5 ~ 19.6 [c] | | | |

[a] Engineering shear strain $\varepsilon_{112}$ corresponding to $\tau_{IS}$, where the slip distances (Å) on the shear plane are in the parentheses.
[b] By Eq. 4 with the input of $\tau_{IS}$ in this Table, $c_{ij}$ in Table 2, and lattice parameter $a_0 = 3.52$ Å for fcc Ni from the present first-principles calculations.
[c] The range of CRSS values for Ni reported in [2,3,54–60]; see details in Supplementary Figure S2.

**Table 2:** Calculated elastic constants (in GPa) of fcc Ni in terms of the conventional cubic lattice ($c_{ij,cub}$) and the orthorhombic lattice ($c'_{ij,orth}$, see Figure 2a for the supercell) without and with pre-strain $\gamma_{110}$.

$c_{ij,cub}$ translated directly from $c'_{ij,orth}$ [a]

$$\begin{pmatrix} 265 & 161 & 161 & 0 & 0 & 0 \\ & 265 & 161 & 0 & 0 & 0 \\ & & 265 & 0 & 0 & 0 \\ & & & 127 & 0 & 0 \\ & & & & 127 & 0 \\ & & & & & 127 \end{pmatrix}$$

$c'_{ij,orth}$ without pre-strain $\gamma_{110} = 0.000$

$$\begin{pmatrix} 340 & 137 & 113 & 0 & 32 & 0 \\ & 340 & 113 & 0 & -32 & 0 \\ & & 365 & 0 & 0 & 0 \\ & & & 79 & 0 & -32 \\ & & & & 79 & 0 \\ & & & & & 101 \end{pmatrix}$$

$c'_{ij,orth}$ with pre-strain $\gamma_{110} = 0.016$

$$\begin{pmatrix} 339 & 138 & 114 & 0 & 32 & -4 \\ & 339 & 114 & 0 & -32 & 0 \\ & & 367 & -8 & 0 & 6 \\ & & & 79 & 6 & -32 \\ & & & & 79 & 0 \\ & & & & & 102 \end{pmatrix}$$

$c'_{ij,orth}$ with pre-strain $\gamma_{110} = 0.033$

$$\begin{pmatrix} 338 & 137 & 113 & 0 & 31 & -9 \\ & 339 & 113 & 0 & -31 & 0 \\ & & 365 & -16 & 0 & 12 \\ & & & 78 & 12 & -31 \\ & & & & 78 & 0 \\ & & & & & 102 \end{pmatrix}$$

$c'_{ij,orth}$ with pre-strain $\gamma_{110} = 0.049$

$$\begin{pmatrix} 341 & 137 & 112 & -1 & 31 & -13 \\ & 341 & 112 & -1 & -31 & 1 \\ & & 365 & -23 & -2 & 17 \\ & & & 76 & 17 & -31 \\ & & & & 76 & 0 \\ & & & & & 102 \end{pmatrix}$$

[a] Experimental elastic constants extrapolated to 0 K [53]: $c_{11} = 261.2$, $c_{12} = 150.8$, and $c_{44} = 131.7$ GPa.

**Table 3:** CPFEM parameter values (see Eq. 6), where $c_{ij}$ are elastic constants of fcc Ni reported in Table 2. All of the parameters in this table were determined through DFT-based calculations in the present study, except $\tau_s$ and $w$. The values for $\tau_s$ were either taken from literature [5] (edge based) or calibrated from macroscopic experiments (edge screw mix), and $w$ was calibrated from macroscopic experiments. See detailed discussion in Section 3.5.

|  | $c_{11}$ (GPa) | $c_{12}$ (GPa) | $c_{44}$ (GPa) | $h_0$ (MPa) | $\tau_0$ (MPa) | $\tau_s$ (MPa) | $w$ |
|---|---|---|---|---|---|---|---|
| Edge based | 265 | 161 | 127 | 24 | 9 | 40 | - |
| Edge screw mix | 265 | 161 | 127 | 120 | 9 | 300 | 0.33 |

**Table 4:** Initial yield stresses (in MPa) from pure Ni single crystal tests by Haasen et al. [2] and Yao et al. [60] together with the corresponding CPFEM predictions in the present work.

| Experimental value | Haasen, $\langle \bar{1}\ 5\ 10 \rangle$ | Haasen, $\langle \bar{1}28 \rangle$ | Yao et al. $\langle 011 \rangle$ |
|---|---|---|---|
|  | 17 | 19 | 31 |
| CPFEM | 21 | 22 | 27 |
| Error compared to experiment | 24% | 18% | 12% |

Supplementary Material for:

# Predictive Crystal Plasticity Modeling of Single Crystal Nickel Based on First-Principles Calculations

John D. Shimanek[1], Shipin Qin[1], Shun-Li Shang[1], Zi-Kui Liu[1], and Allison M. Beese[1,2,*]

[1] *Department of Materials Science and Engineering, The Pennsylvania State University, University Park, PA 16802, USA*

[2] *Department of Mechanical Engineering, The Pennsylvania State University, University Park, PA 16802, USA*

\* *Corresponding author: amb961@psu.edu*

## 1. Details of first-principles calculations

All DFT-based first-principles calculations in the present work were performed by the Vienna *Ab initio* Simulation Package (VASP) [1]. The ion-electron interaction was described by the projector augmented wave (PAW) method [2]; the exchange-correlation functional was characterized by the generalized gradient approximation (GGA, PW91) as parameterized by Perdew et al. [3]; and the core configuration of [Ar] was employed for Ni, as recommended by VASP. In VASP calculations, the *k*-point meshes of 10×16×7 were used for the 6-atom orthorhombic supercell (see Figure 2a in the main text); the cutoff energy of 337 eV (i.e., the precision of "high" used in VASP) was employed for the plane-wave basis set; and the energy convergence criterion of electronic self-consistency was selected as $10^{-5}$ eV per supercell for all calculations. The reciprocal-space energy integration was performed by the Methfessel-Paxton [4] technique with a 0.2 eV smearing width, which can result in accurate total energies as well as stresses. Concerning pure alias shear deformation, an external optimizer GADGET developed by Bučko et al. [5] was used to control both stresses and forces acting on each atom during VASP calculations. The relaxed stresses (except for the shear stresses due to the fixed $\gamma_{112}$ and/or $\gamma_{110}$ values) were less than 0.15 GPa, and the forces acting on atoms were less than 0.03 eV/Å. Spin polarization was considered in all first-principles calculations due to the magnetic nature of Ni.

To explore the layer dependency of ideal shear strength, ancillary DFT-based calculations of pure alias shear along $\{111\}\langle11\bar{2}\rangle$ were also performed using the 6-atom (3-layer), 12-atom (6-layer), and 18-atom (9-layer) orthorhombic supercells based on the structure shown in Figure 2a. The corresponding *k*-point meshes were 10×16×7, 9×16×3, and 7×12×2, respectively. In addition, phonon calculations were also carried out to explore the origin of layer-dependent $\tau_{IS}$ in terms of the 6-atom (3-layer) and the 12-atom (6-layer) orthorhombic cells after $\{111\}\langle11\bar{2}\rangle$ pure alias shear by applying the same amount of shear displacement (0.5 Å). These phonon calculations were performed by the supercell approach [6] as implemented in the YPHON code [7,8]. The VASP code was again the computational engine in calculating force constants by the density functional perturbation theory. For both the 3-layer and the 6-layer orthorhombic lattices, the 72-atom supercells together with the 3×3×2 $k$-point meshes were used for phonon calculations. Note that all other conditions used for these ancillary first-principles calculations were the same as the aforementioned settings.

The ideal shear strength of the orthorhombic supercell depended on the number of layers (instead of the number of atoms per layer) in the [111] direction, with more layers resulting in lower ideal shear strengths. Table S1 summarizes this relation by showing the predicted ideal shear strengths of fcc Ni by pure alias shear deformation along $\{111\}\langle11\bar{2}\rangle$ without any pre-strain for 3, 6, and 9 layers of atoms along [111]. The 3-layer, 6-atom model produces the maximum shear strength of 5.15 GPa, which agrees well with previous predictions of around 5.0 GPa using pure alias or pure affine shear deformations [9]. The current value also agrees reasonably well with the values estimated from nanoindentation of approximately 8 ± 1.5 GPa [10] or 6.4 ± 1.1 GPa [11]. The higher values found by nanoindentation are likely due to the measurement being performed on a non-close packed (001) plane [10] and the stabilizing effect of the triaxial stress state beneath the indenter tip [12]. With an increasing number of {111} layers, the predicted $\tau_{IS}$ decreased significantly despite the fact that the absolute displacement distance increased only slightly. The 3-layer, 6-atom supercell was chosen for study in the present work due to its agreement with experimental estimates of the ideal shear strength of pure Ni and because it represents the minimum number of layers, and therefore maximum shear stress.

To understand the decrease of $\tau_{IS}$ with increasing numbers of {111} layers, the stretching force constants are plotted in Figure S1 with phonon calculations for two fcc-based orthorhombic lattices: one with 3 layers (6 atoms) and one with 6 layers (12 atoms) after pure alias shear with the same amount of displacement distance (0.5 Å) applied. Here the force constants, particularly the dominant stretching force constants shown in Figure S1 (as opposed to the significantly smaller bending force constants), provide quantitative understanding of the interaction or bonding between atomic pairs [13,14]. A large and positive force constant indicates strong bonding, while a negative force constant suggests the pair of atoms tend to separate from each other. Figure S1 shows that the maximum stretching force constants from the 3-layer lattice are higher than those from the 6-layer lattice (2.46 versus 2.28 eV/Å$^2$), indicating the bonding between atoms becomes weaker with an increasing number of {111} layers during pure alias shear deformation, which results in lower $\tau_{IS}$ values.

It should be noted that all DFT-based calculations of CRSS in the present work were performed at 0 K for simplification, while all experimental data were taken at room temperature. This simplification is appropriate because, for pure metals, the CRSS values at 0 K are close to those at room temperature [15]. Additionally, previous calculations have indicated that properties from DFT-based calculations at 0 K are comparable to experimental data measured at room temperature (298 K) for many properties. For example, the predicted difference of enthalpy of formation is negligible between 0 K and room temperature (< 0.2 kJ/mol for metal sulfides [16]), the predicted bulk moduli of Ni and Ni$_3$Al decrease about 9 GPa (5 %) from 0 K to room temperature [17], and the predicted ideal shear strength of Ni decreases about 0.1 GPa (2 %) [9].

Lastly, the conversion of the DFT-based ideal shear strengths and elastic constants depends on the choice of elastic factor, whose value depends on dislocation character. These elastic factors have been derived for an anisotropic solid by Hirth and Lothe [18]. For example, for an edge dislocation aligned with the z-direction, with a Burgers vector $\boldsymbol{b} = (b_x, b_y, 0)$, the corresponding $K_{e_x}$ of edge dislocation along the *x*-direction is given by [18],

$$K_{e_x} = (\bar{c}'_{11} + c'_{12}) \left[ \frac{c'_{66}(\bar{c}'_{11} - c'_{12})}{(\bar{c}'_{11} + c'_{12} + 2c'_{66})c'_{22}} \right]^{1/2} \qquad \text{Eq. S1}$$

where $\bar{c}'_{11} = (c'_{11} c'_{22})^{1/2}$ and $c'_{ij}$ indicates the transformed elastic constants onto the slip system of interest. In the present work, the transformed lattice vectors of fcc Ni are parallel to the $[11\bar{2}]$, $[\bar{1}10]$, and $[111]$ directions of the conventional fcc lattice, i.e., the $\boldsymbol{a}_{\text{orth}}$ (*x*), $\boldsymbol{b}_{\text{orth}}$ (*y*), and $\boldsymbol{c}_{\text{orth}}$ (*z*) directions, respectively; see Figure 2a. Notably $K_{e_x} = K_{e_y} (= K_e)$ for edge dislocations along the *x*- and *y*-directions for the present fcc Ni represented by the orthorhombic cell as shown in Figure 2a. The elastic factor for screw dislocations, $K_s$, of an anisotropic crystal is given by [18],

$$K_s = [c'_{44} c'_{55} - (c'_{45})^2]^{1/2} \qquad \text{Eq. S2}$$

## 2. Interpretation of experimental data in the literature

In the works of Yao et al. [19] and Haasen [20], both of which were used for comparison purposes in Section 3.3 and beyond, the authors showed only the resolved shear stress and resolved shear strain data. However, it is not straightforward to convert directly measurable quantities in the tests, namely force and displacement, to resolved shear stress and resolved shear strain on slip systems; the conversion process depends on the assumptions made as discussed below [21].

In the work by Yao et al. [19], only one slip system was assumed to be operating. The resolved shear strain $\gamma$ and the resolved shear stress $\tau$ under this assumption are calculated as [22–24]:

$$\gamma = \frac{1}{\cos\theta_0}\left[\sqrt{(1+\varepsilon^{eng})^2 - \sin^2\lambda_0} - \cos\lambda_0\right] \qquad \text{Eq. S3}$$

$$\tau = \sigma^{eng}\frac{\cos\theta_0}{1+\varepsilon^{eng}}\sqrt{(1+\varepsilon^{eng})^2 - \sin^2\lambda_0} \qquad \text{Eq. S4}$$

where $\theta_0$ is the initial angle between the loading direction and the slip plane normal direction, $\lambda_0$ is the initial angle between the loading direction and the slip direction, $\sigma^{eng}$ is the engineering stress, and $\varepsilon^{eng}$ is the engineering strain. This approximation assumes that the loading axis continually rotates with respect to the active slip system throughout loading, which is unlikely to be true in finite deformation [25]. Eqs. S3 and S4 were used to calculate the engineering stress-strain curve in the tests in ref. [19].

In the framework of double slip, the rotation of the loading axis with respect to the active slip system is assumed to cease when it reaches a specific orientation. Before reaching this orientation, single slip operates, and the equations above can be applied. After the rotation of the loading axis activates a conjugate slip system, the two slip systems are assumed to operate simultaneously with the same hardening rate, rotating the loading axis along the slip system boundary until reaching a point of stable double glide that prevents further rotation [22]. If $\boldsymbol{n}_1$ and $\boldsymbol{n}_2$ are the unit normals of the two slip planes, and $\boldsymbol{u}_1$ and $\boldsymbol{u}_2$ are the unit vectors of the two slip directions, the resolved shear strain $\gamma$ and resolved shear stress $\tau$ under the double glide approximation can be calculated as [22,26]:

$$\gamma = \frac{2}{\boldsymbol{n}_1\boldsymbol{u}_2}\ln\left[1 + \frac{\boldsymbol{n}_1\boldsymbol{u}_2}{|\boldsymbol{v}|}\frac{\sin\beta_0}{\cos\theta_0}(\cot\beta - \cot\beta_0)\right] \qquad \text{Eq. S5}$$

$$\tau = \sigma^{eng}\frac{|\boldsymbol{v}|}{2}\cos\beta\left\{\cos\theta_0 + \frac{\boldsymbol{n}_1\boldsymbol{u}_2}{|\boldsymbol{v}|}\sin\beta_0(\cot\beta - \cot\beta_0)\right\} \qquad \text{Eq. S6}$$

$$\sin\beta = \frac{\sin\beta_0}{1+\varepsilon^{eng}} \qquad \text{Eq. S7}$$

where $\boldsymbol{v} = \boldsymbol{u}_1 + \boldsymbol{u}_2$, and $\beta_0$ is the angle between the loading direction and $\boldsymbol{v}$ at the onset of double glide. Eq. S5 through S7 were adopted in the present work to calculate the engineering stress-strain curves in Haasen's tests, in which the initial loading direction was $\langle\bar{1}\ 5\ 10\rangle$ for crystal #6 and $\langle\bar{1}28\rangle$ for crystal #18 in ref. [20]. In both tests, the $\{111\}\langle\bar{1}01\rangle$ slip system was active first. It was assumed that when the loading direction rotated to $\langle\bar{5}\ 5\ 14\rangle$ for crystal #6 and to $\langle\bar{2}29\rangle$ for crystal #18, double slip began and $\{\bar{1}\bar{1}1\}\langle 011\rangle$ started to operate as an additional slip system. The engineering stress-strain curves for all three tests, calculated using the above equations [19,20], are shown in Figure 4b of the main text.

# Figures

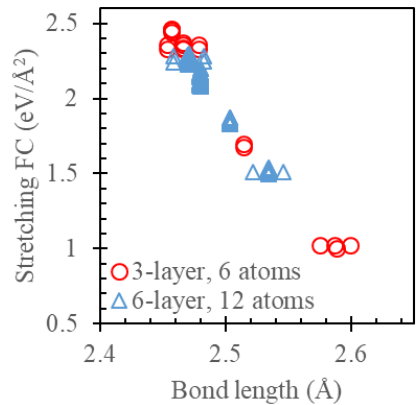

**Figure S1:** Stretching force constants (FCs) as a function of bond length for two fcc lattices of Ni: (i) the orthorhombic lattice with 3 layers and 6 atoms (see Figure 2), and (ii) the orthorhombic lattice with 6 layers and 12 atoms. Note that both lattices have the same shear displacement of 0.5 Å for the $\{111\}\langle11\bar{2}\rangle$ shear deformation, and the 72-atom supercells were employed for phonon calculations of both lattices.

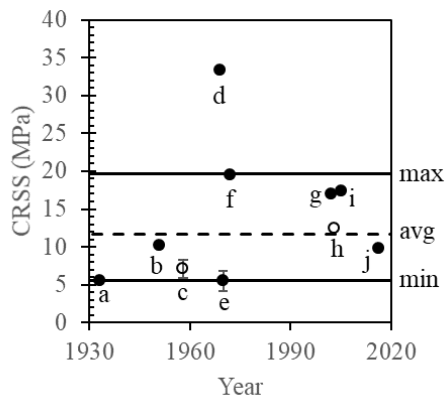

a: Osswald (99.7 wt.% Ni) [28],
b: Andrade et al. (99.9 wt.% Ni) [29],
c: Haasen (99.999 wt.% Ni) [20],
d: Latanision et al. (99.8 wt.% Ni) [27],
e: Venkatesan et al. (unknown purity) [30],
f: KondratEv et al. (99.999 wt.% Ni) [31],
g: Hecker et al. (99.99 wt.% Ni) [32],
h: Yao et al. (99.999 wt.% Ni) [19],
i: Dimiduk et al., (unknown purity) [33],
j: Luo et al. (99.99 wt.% Ni) [34].

**Figure S2:** Initial CRSS values of pure Ni reported in the literature. The value reported in ref. [27] was significantly higher than others and was excluded from the present study. The open symbols (c and h) correspond to the studies adopted in the present study for validation of CPFEM predictions.

**Tables**

**Table S1:** Ideal shear strength ($\tau_{IS}$), associated slip (displacement) distance on the shear plane, and engineering shear strain $\gamma_{112}$ of fcc Ni due to pure alias shear along $\{111\}\langle11\bar{2}\rangle$ using supercells with different layers, with the total number of atoms within each supercell given.

| Supercell | Slip distance (Å) | Shear strain $\gamma_{112}$ | $\tau_{IS}$ (GPa) |
|---|---|---|---|
| 3-layer (6 atoms) | 0.78 | 0.13 | 5.15 |
| 6-layer (12 atoms) | 0.80 | 0.07 | 3.61 |
| 9-layer (18 atoms) | 0.80 | 0.04 | 2.60 |

**Table S2**: Statistics of the initial CRSS values in Fig. S2. The outlier reported in ref. [27] is excluded.

| Max | Min | Average | Std. Dev. | Relative error |
|---|---|---|---|---|
| 19.6 MPa | 5.5 MPa | 11.66 MPa | 5.01 MPa | 43% |